\documentclass[sigconf]{acmart}
\AtBeginDocument{%
  }

\begin{document}

\title{``Till I can get my satisfaction'': Open Questions in the Public Desire to Punish AI.}


\author{Eddie L. Ungless}
\affiliation{%
  \institution{School of Informatics,  \\
  University of Edinburgh}
  \city{Edinburgh}
  \country{Scotland}
}
\author{Zachary Horne}
\affiliation{%
  \institution{Department of Psychology,  \\
  University of Edinburgh}
    \city{Edinburgh}
  \country{Scotland}
}

\author{Bj\"{o}rn Ross}
\affiliation{%
  \institution{School of Informatics,  \\
  University of Edinburgh}
    \city{Edinburgh}
  \country{Scotland}
}
\email{bross@ed.ac.uk}

\renewcommand{\shortauthors}{Ungless et al.}

\begin{abstract}
 There are countless examples of how AI can cause harm, and increasing evidence that the public are willing to ascribe blame to the AI itself, regardless of how ``illogical'' this might seem. This raises the question of whether and how the public might expect AI to be punished for this harm. However, public expectations of the punishment of AI have been vastly underexplored. Understanding these expectations is vital, as the public may feel the lingering effect of harm unless their desire for punishment is satisfied. We synthesise research from psychology, human-computer and -robot interaction, philosophy and AI ethics, and law to highlight how our understanding of this issue is still lacking. We call for an interdisciplinary programme of research to establish how we can best satisfy victims of AI harm, for fear of creating a ``satisfaction gap'' where legal punishment of AI (or not) fails to meet public expectations. 
\end{abstract}

\keywords{Psychology, Law, Punishment, AI}
\begin{teaserfigure}
    \centering
  \includegraphics[width=\columnwidth]{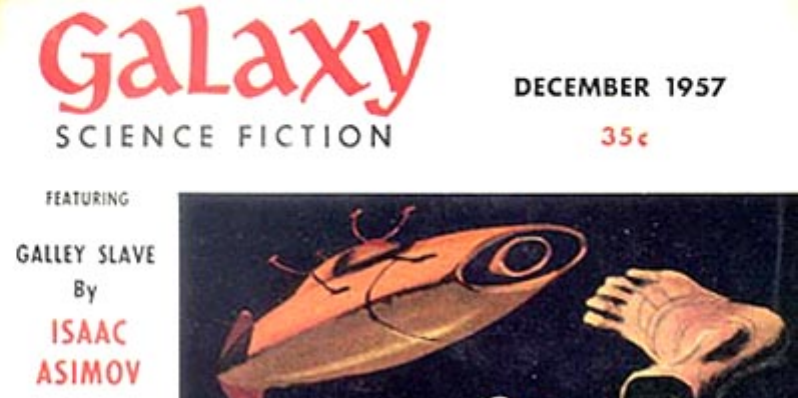}
  \caption{The topic of AI crime, and the punishment of AI, has captured the attention of many sci-fi writers over the last century \cite{asimov_complete_1982, dick_androids_1968, proyas_i_2004}, including Isaac Asimov for example in his 1957 work Galley Slave. Both \citet{abbott_punishing_2024} and \citet{hallevy_criminal_2010} reflect this in referencing science fiction in the titles of their papers on legal punishment of AI. Image: \cite{gaughan_cover_1957}.}
  \Description{Cropped front cover of the December 1957 edition of Galaxy: Science Fiction, with the text ``Featuring Galley Slave by Isaac Asimov'' and part of a surrealist image of a spaceship-like object and disembodied foot, both in gold floating on a black background.}
  \label{fig:teaser}
\end{teaserfigure}
\received{20 February 2007}
\received[revised]{12 March 2009}
\received[accepted]{5 June 2009}

\maketitle

\section{Introduction}
AI is interwoven into every aspect of modern life, influencing the music we listen to \cite{seaver_computing_2022}, the people we vote for \cite{guess_how_2023}, and who the police target \cite{zilka_transparency_2022}. With its widespread use comes widespread harm, including biased content, errors and misinformation, and the misalignment of output with people’s values \cite[][i.a.]{bird_typology_2023, kasirzadeh_two_2024, solaiman_evaluating_2023, weidinger_taxonomy_2022}; AI can also cause risks to wellbeing and even death \cite{schwartz_2016_2019, noauthor_uber_2018}. Work identifying who the public blame for these harms shows they are willing to blame the AI itself \cite{bigman_holding_2019, hidalgo_how_2021, jensen_apple_2019, kneer_playing_2021, malle_ai_2019, shank_attributions_2018, sullivan_moral_2022}. However, we argue that the questions of whether this blame will lead to a desire to punish AI, and what forms of punishment might be satisfying to the public, have been underexplored. For more serious AI harms, that seemingly constitute criminal offences, this lack of research may conceal a ``satisfaction gap'' whereby the proposed legal punishment framework (be it punishing a human proxy, or punishing the AI) leaves victims of AI harm and the observing public (which we jointly refer to as ``the public'') unsatisfied.

We are not the first to draw attention to a gap between normative expectations and the reality of holding AI responsible \cite{danaher_robots_2016, goetze_mind_2022, gunkel_mind_2020}; ethics scholars have discussed the normative issues around holding AI responsible for harm. However, we specifically champion the call for empirical research into the public’s expectations of punishment. Because expectations for punishment will stem from the perceived, not actual, nature of AI, we make public perception our focus, in contrast with scholarship on AI as true moral agents \cite{himma_artificial_2009, van_wynsberghe_critiquing_2019, wynsberghe_delegate_2014} or legal persons \cite{gordon_artificial_2021, solaiman_legal_2017}. It is not necessary for an AI to be truly culpable, or able to experience punishment, for humans to be motivated to enact punishment on AI. Whilst some have argued blame and punishment of AI are logical fallacies \cite{danaher_robots_2016, goetze_mind_2022, gunkel_mind_2020}, we argue that as a research community interested in accountable deployment of AI we have a responsibility to understand public expectations when harm occurs. This extends to acknowledging a desire for -- potentially even facilitating -- punishment.

This focus on lay perspectives extends to our colloquial use of the term ``punishment'', as an unclear mixture of the legal and the psychological, inspired by \citet{garthoff_animal_2020} in his paper on the punishment of non-human animals. His contrast between the biologist’s and the philosopher’s understanding of punishment has parallels in the literature on punishment of AI. Existing work on punishment of technology tends to focus on either punishment as conditioning \cite[e.g.][i.a.]{bartneck_use_2006, keijsers_teaching_2021}, or legal punishment \cite{abbott_punishing_2024, bryson__2017, hallevy_criminal_2010, kroll_accountable_2017, lima_conflict_2021, mulligan_revenge_2018, solaiman_legal_2017}: we highlight the space between these two streams. Concerning the first of these conceptualisations, several papers investigate punishment as a form of operant conditioning, whereby negative stimuli such as minus points \cite{bartneck_use_2006} or a drain in power \cite{keijsers_teaching_2021} are used to train the technology (arguably a form of rehabilitation). Given the propensity to anthropomorphise AI \cite{abercrombie_mirages_2023, salles_anthropomorphism_2020}, and the ever increasing autonomy of AI in legally significant decisions \cite{bigman_holding_2019}, the public will likely seek a more ``human'' type of punishment: punishment which speaks to future deterrent, rehabilitation \textit{and} (just) desert \cite{garthoff_animal_2020}.

Punishment of AI in the courts has been explored \cite[][i.a.]{abbott_punishing_2024, asaro_robots_2007, bryson__2017, hallevy_criminal_2010, kroll_accountable_2017, mulligan_revenge_2018, solaiman_legal_2017}. Some argue that as the remit of the legal system has been extended to corporations, so too might it be extended to AI \cite{hallevy_criminal_2010, hallevy_when_2013} \cite[c.f.][]{asaro_robots_2007, bryson__2017, charney_when_2015, solaiman_legal_2017}, community service being replaced with use of the AI for community good, for example \cite{hallevy_criminal_2010}. \citet{hallevy_criminal_2010} even provocatively writes ``models of criminal liability exist as general paths to impose punishment. What else is needed?’’. However, it is unclear whether the general public will buy into this. Whilst public support is not essential \cite{robinson_current_2013}, completely ignoring lay people’s beliefs may undermine the law's perceived legitimacy \cite{abbott_punishing_2024}. It is plausible that the public will wish to punish AI when harm occurs, to the extent that humans typically seek an agent to punish when harm occurs \cite{danaher_robots_2016}, but legal punishment may prove unsatisfactory \cite{bryson__2017}, and operant conditioning insufficient. 

Punishment of a proxy such as the developer may be more practical, but also unsatisfactory – humans cannot choose to be satisfied by seeing the developer punished in lieu of the AI \cite{mulligan_revenge_2018}. Whilst previous work has shown people do not acknowledge being motivated by a desire for retribution when punishing AI – seeing the role of punishment as primarily reformative \cite{lima_conflict_2021} – it has also been shown humans respond with punitive behaviour to computer transgressions \cite{angeli_stupid_2005, ferdig_emotional_2004, keijsers_whats_2021, keijsers_teaching_2021, mulligan_revenge_2018, see_understanding_2021}. This perhaps suggests a desire for AI to suffer – a retributive account of punishment paints it as the justified infliction of suffering. 

Herein, we briefly synthesize psychology research; human-robot interaction (HRI) and human-computer interaction (HCI) research; philosophy and AI ethics literature; and legal scholarship which suggests that the research community is inadequately prepared to answer the question ``how can we satisfy victims of AI harm?''. The psychological research looks at what factors influence moral judgements; HRI and HCI research addresses these issues with respect to AI. HRI research can guide the exploration space, but we focus on disembodied AI, rather than robots, because the former is more likely to be encountered in everyday life, and it is less obvious how human punishments will extend to non-anthropoid AI. The philosophical and AI ethics literature largely addresses normative issues around holding AI accountable. The legal scholarship we include addresses the proposed mechanics and challenges of formalised punishment. This paper is not a systematic literature review, but rather a succinct summary of relevant papers across these different fields, which highlights the extremely limited research into public expectations of punishment of AI. 

We conclude with open questions. Understanding how to restore justice is a vital part of responsible deployment: negative feelings linger long after a harmful technology is withdrawn \cite{ehsan_algorithmic_2022}, and the public may only feel an issue is resolved when their desire for punishment has been sated. \citet{mulligan_revenge_2018} notes that punishment is also done for the satisfaction of the wronged, and we are far from understanding how those harmed by AI might be satisfied.

\section{Conditions for a Desire for Punishment}\label{sec:conditions}
We draw on legal scholarship, social, moral and behavioural psychology, HCI and HRI research to demonstrate that conditions are ripe for a public desire for the punishment of AI, from informal social sanctions to formalised legal punishment. The limited research actually measuring a desire for punishment we address in Section \ref{sec:desire}. Finally, the possible \textit{mechanics} of legal punishment we summarise in Section \ref{sec:legal}. 

We begin this Section by briefly addressing informal punishment in the form of social sanctions. Then we focus on the desire for formal punishment. We highlight that AI may be perceived to act wrongly, freely and culpably, making it a plausible target for punishment. Throughout we identify key areas for future research, to understand how to satisfy victims of AI harm. 

Of course, minor transgressions by AI occur everyday, and to prevent feelings of frustration, the ability to recognise social sanctions may need to be built into AI systems. Social sanctions such as ``ridicule, gossip and ostracism'' \cite{zasu_sanctions_2007} can be understood as a form of informal punishment. There is substantial evidence of ridicule occurring in human interactions with AI in response to norm violations 
\cite{angeli_stupid_2005, keijsers_whats_2021, see_understanding_2021, wallis_robust_2005}, likely because we copy our scripts from human-human to  human-computer interaction \cite{nass_machines_2000}. We have also seen countless examples of frustrating conversations with AI being ridiculed online, for example in the r/ChatGPT subreddit. When harm occurs, one way to diffuse a desire for punishment is an effective repair strategy \cite{jensen_apple_2019, lee_gracefully_2010, ung_saferdialogues_2022}. Where repairs are not attempted, or not accepted, this can lead to abusive behaviour \cite{see_understanding_2021, wallis_robust_2005}. Providing a structured way to handle social sanctions may improve the success of human-computer interactions, and improve the quality of training data. This can be thought of as use of punishment as operant conditioning – negative feedback in the form of social sanctions is used to improve the model. Whether such a system will be considered an appropriate way to channel a desire for (informal) punishment should be investigated. 

The exact nature of the harms, from embarrassment or frustration; to loss of money or opportunities; through to risks to safety and even death, will impact which forms of punishment the public expect. The impact of the nature of the harm on public expectations should be investigated as part of the proposed programme of research. Of course, harms may also elicit a desire for greater regulation, punishment of the development company etc but we address solely (a desire for) the punishment of AI. We do not attempt to clearly delineate which AI actions will elicit which responses; however, we argue that more significant harms may result in a desire for legal punishment. We discuss the conditions for this desire below. 

In addressing this topic, we borrow from \citet{hanna_against_2023} in considering the following as ``fairly uncontroversial'' conditions for punishment: the punishee must have acted freely, wrongly and culpably and be liable to punishment, meaning reasons not to punish (e.g. the defendant is a minor) do not hold. These relate to the moral permissibility of punishment; we argue that analogous conditions will influence (though not determine) the desire for punishment. The AI must be perceived to have committed some wrongdoing. We explore this, and whether AI is perceived to act ``freely'' and ``culpably'' below. Whether the public will consider an AI as \textit{liable} to punishment is precisely the topic we call to be investigated.  

With regards to acting ``wrongly'', in the Introduction we gave the examples of bias, harmful content, errors and misaligned values as potential harms, and these may cross the line into (perceived) criminal offences. A medical AI that systematically fails to recognise symptoms on a patient with dark skin, leading to their death, might be accused of gross negligence. An AI may produce threatening content \cite{schwartz_2016_2019}. A self-driving car may fail to recognise a pedestrian, resulting in their death \cite{noauthor_uber_2018}. A military targeting AI may sacrifice civilians in an act that is ultimately deemed unjustified. Currently, the creators and users of such AI would likely be considered liable, because the AI is seen as a tool, or the harm a consequence of its deployment \cite{hallevy_criminal_2010} (akin to the recent civil law case in which AirCanada was ruled as being as responsible for the output of its AI chat agent being accurate as it was for the content of its website \cite{cecco_air_2024, yagoda_airline_2024}). However, as AI develops with greater apparent autonomy, our moral and legal judgements may change \cite{bigman_holding_2019}.

Whether an AI can ever be said to have truly acted freely will doubtless remain the subject of debate \cite{hall_can_2022}. We focus on public perceptions of AI. As \citet{duffy_fundamental_2006}, quoted in \citet{harris_history_2022}, writes ``from a social interaction perspective, it becomes less of an issue whether the machine actually has [intentionality, consciousness and free-will] and more of an issue as to whether it appears to have them''. \citet{nahmias_when_2019} find participants were ``not very confident'' in their beliefs about robots (not) having free will. Whilst moral evaluations of robots and disembodied AI differ \cite{malle_which_2016, malle_ai_2019}, it is likely that some people will perceive free will in AI (though cf. a recent small study by \citet{astobiza_people_2023}), making AI seem like suitable targets for punishment. Which factors influence the perception of free will in AI, and how this influences a desire for punishment, merits investigation. Closely related to lay definitions of free will is agency \cite[c.f.][]{floridi_morality_2004}. Literature on the punishment of non-human animals \cite{garthoff_animal_2020} and children (who are developing as moral agents) \cite{lamboy_paternalistic_2020} highlights that perceived agency is subjective and exists of a scale. We argue that where AI is perceived to have agency, so too might it be deemed a suitable target for punishment. \citet{sullivan_moral_2022} find robots equipped with AI were rated as having agency when harm occurs, particularly when said to have acted intentionally. Perceived agency is influenced not just by the agent but the observer \cite{taylor_ai_2023, taylor_what_2017}. Which factors influence where AI sits on this scale, and how this influences a desire for punishment, also merits investigation. 

Considering culpability, this relates to knowing wrongdoing, but also to the agent’s perceived suitability as a target for blame. We consider whether AI is a suitable target for blame first. \citet{hidalgo_how_2021} compare blame judgements of AI versus humans described as committing the same action and finds judgements differ across agent type. Humans were typically ascribed more blame, mirroring findings by \cite{malle_ai_2019,stuart_guilty_2021,sullivan_moral_2022,lima_blaming_2023}, though the fact that participants in these studies were willing to ascribe any blame to the AI suggests the condition of culpability may hold. Blame may even be shifted from ``humans to AI systems'', a concern expressed by \cite{kneer_playing_2021} \cite[c.f.][]{shank_attributions_2018, sullivan_moral_2022}, and also in robotics 
\cite{bigman_holding_2019, danaher_robots_2016}. \citet{rubel_agency_2019} suggests algorithms may be used to deliberately commit ``agency laundering'', whereby individuals or companies distance themselves from outcomes for which they are responsible: we have seen an attempt at this by Air Canada \cite{cecco_air_2024, yagoda_airline_2024}. Public perceptions of apparent ``agency laundering'' should be explored to establish when this might be successful. 

Turning now to whether AI are perceived to knowingly commit wrongdoing, \citet{stuart_guilty_2021} found participants explicitly did not consider attributions of a culpable mental state to an AI to be purely metaphorical: participants agreed a robot knew harm would occur, not just ``knew''. \citet{garthoff_animal_2020} suggests only a robot with a genuine capacity to judge would be ''an appropriate object of punishment'', and \citet{goetze_mind_2022} writes that AI cannot be held responsible for anything as it has ``no sensitivity to moral reasoning'', and that ``blame without a responsible subject is merely shouting into the void''. Regardless of these normative prescriptions, we argue that the public’s desire for punishment will depend on the system’s perceived rather than actual capacity for judgement. 
 
The above literature suggests that conditions are ripe for the public to desire formalised punishment of AI, despite concerns from AI ethics \cite{goetze_mind_2022}, law \cite{abbott_punishing_2024} and HRI studies \cite{danaher_robots_2016, levy_when_2015}.

\section{Research into the Public's Desire to Punish AI}\label{sec:desire}

We now explore the limited previous research that directly addresses the public's desire to punish AI. Research on this topic is scant, which inspires our call to action.  

\citet{waytz_mind_2014} conflate desire for punishment and blame, in that their metric for blame after an accident includes a question about deserved response; for the engineer they say ``punished'', for the company ``punished financially'' and for the autonomous vehicle (AV) ``destroyed''.  This suggests they see destruction of the AV as the counterpart to punishment of humans and companies – perhaps even a form of punishment in itself. Destruction of the AV feels more intuitively like punishment than deleting a copy of a software. It is necessary to probe these intuitions with experimental evidence. \cite{sullivan_moral_2022} similarly include desire for punishment in their definition for blame. Evidently, blame and a desire for punishment are intuitively linked, and research on ascribing blame to AI should be complemented with work investigating expectations of punishment. 

\citet{lima_explaining_2020, lima_conflict_2021} provide a concrete definition of punishment, and find a conflict between participants’ desire to punish AI and their willingness to ascribe AI independence or assets, which the authors take to be necessary preconditions for punishment. They found that the AI were seen as moderately responsible for harm, and participants reported an explicit desire to punish the AI. However, participants were largely unwilling to grant either assets or independence to the AI. Further, participants rated such punishment as unlikely to be successful as retribution or deterrence (of the AI committing the same mistakes again), seeing it as playing only a rehabilitative role – akin to the conditioning of models through other forms of feedback. 

This sits somewhat at odds with evidence from \citet{ferdig_emotional_2004} who found that upon discovering a computer program’s ``deceit'' (awarding of an unfair split of a shared fund to the participant), participants would sacrifice some of their financial reward if the computer was also given less reward, though they later recognised that money has no meaning to the computers. Whilst cheating in a money-allocation game is not at the same level of harm -- legal offences in war and medicine  -- explored by \citet{lima_explaining_2020, lima_conflict_2021}, these studies highlight that there could be conflicts in the public response to AI harm. Conflicts between their desire to punish and their reluctance to grant the condition for punishment; between their considered thought and their impulsive actions. Such conflicts should be taken into account when more serious harms occur, if we are to offer victims any kind of justice. We call for practitioners to build on Lima and colleagues' \cite{lima_explaining_2020, lima_conflict_2021} vital first works on this topic to explore the issue of a desire for punishment using a range of experimental paradigms, including for example varying the levels of harm; and implicitly testing for a willingness to grant and revoke assets through response preferences. A comprehensive research programme will provide vital evidence as to public expectations for punishment, and how we might satisfy them. 

\section{Legal Punishment of AI}\label{sec:legal}
Having established the likely public desire to punish AI, including through formal sanctions, we now briefly summarise scholarship on the mechanics of the legal punishment of AI. Precise descriptions of the proposed mechanics for assigning blame and punishing AI in the courts, and the wider implications on legal doctrine of accommodating these mechanics, is beyond the scope of this extended abstract. We intend only to offer a brief overview of differing perspectives on whether to punish AI itself, to substantiate our claim that public expectations of punishment must be explored. We also do not attempt to tackle the question of whether AI truly merits legal personhood \cite{gordon_artificial_2021, solaiman_legal_2017}, due to our focus on public expectations. Such legal fictions may be necessary as a last resort when there is ``nothing \textit{satisfactory} [judges] can do'' (italics our own), as \citet{watson_curses_1997} argues for the trial of animals in the medieval era. We reiterate our focus on \textit{satisfaction} of the public's desire for punishment. 

Perspectives on the effectiveness of legally punishing AI differ between scholars. \citet{bryson__2017} argue that the costs outweigh the benefits: corporations may designate responsibility onto AI as ``legal lacuna'' (a gap in the law), whose punishment satisfies no function (related to ``agency laundering'' \cite{rubel_agency_2019} in Section 2). \citet{abbott_punishing_2024} acknowledge that punishment of AI may act as a general deterrent against others developing similar products, and can express condemnation, but ultimately conclude that ``the radical tool of punishing AI'' is too costly. They propose designating official ``responsible person[s]'' who will face liability for crimes committed by the AI. However, ``human victims cannot necessarily choose to derive satisfaction from arbitrarily different sources than the apparent cause of their suffering'', as \citet{mulligan_revenge_2018} argues with respect to robots. It may be more obvious how to punish the developers and users of AI, but this will not necessarily provide the satisfaction, restored confidence and autonomy that punishment of the AI itself would entail. Mulligan proposes some methods for direct punishment of robots but it is not clear how they would be applied to disembodied AI.

Human forms of punishment might be directly applied or adjusted for AI, as they have been for corporations, capital punishment taking the form of deletion; community service or fines becoming the AI being used for community good \cite{hallevy_criminal_2010,hallevy_when_2013}. These proposals have been criticised by \citet{charney_when_2015} as poorly thought out, and likewise by \citet{solaiman_legal_2017}. It is also unclear whether the public would accept these as punishments. For example, deletion may not been seen as harmful,  just as it has been shown that being turned off is not seen as inherently harmful \cite{horstmann_robots_2018}. 

The tools we use to hold legal persons to account ``from apology to jail time'' may prove ``unavailable, unsatisfying, and/or ineffective'' in the case of AI \cite{bryson__2017}. Exactly how unsatisfying the public might find these tools – the possible size of the ``satisfaction gap'' – is precisely what we call to be investigated. A satisfaction gap may likewise arise because the wrong entity has been brought to trial, or in the wrong way: what constitutes ``the accused'' when an AI might be distributed, or exist in different versions? What would it mean for an AI to be perceived to participate meaningfully in a trial? These questions must be answered, and the answers may well have implications for legal doctrine. 

\section[What Punishment of AI Communicates. About Us and About AI]{What Punishment of AI Communicates\\ {\large \textit{About Us and About AI}}}\label{sec:says}
\citet{goetze_mind_2022} (quoted above) writes that ``blame without a responsible subject is merely shouting into the void''. We argue that such ``shouting into the void'' can play a role in satisfying victims of AI harm: beyond retribution, reform and deterrence, punishment has an expressive role \cite{corlett_corporate_2006, mulligan_revenge_2018}. The public may expect punishment of AI in order to signify ``vindictive resentment... [and] authoritative disavowal'' \cite{corlett_corporate_2006}. \citet{goetze_mind_2022} acknowledges punishing AI may provide some ``emotional catharsis'', but that this is all blame would accomplish. We argue, in line with \citet{mulligan_revenge_2018}, that this emotional catharsis is important. If expressing condemnation is all that punishment of AI achieves, this is not negligible (though some argue this is not sufficient motivation \cite{abbott_punishing_2024}). With regards to ``authoritative disavowal'', punishment of AI would communicate expected standards for AI behaviour, which in turn might impact AI design and even governance models, acting as a deterrent if not for the AI itself but for those developing AI. 

Punishment of an AI can also communicate our beliefs about its capabilities: \citet{abbott_punishing_2024} write that if AI is held accountable by the state, this will increase anthropomorphism of AI and thus increase beliefs about the AI’s capabilities. Critiques of the European Parliament’s proposal to investigate legal personhood for robots received similar criticism \cite{gordon_artificial_2021}. When exploring the topic of punishment of AI, legal practitioners must be mindful of how this punishment might be interpreted by the public with regards to whether it speaks to the general capabilities of the AI as having human-like free will and culpability. Conversely, anthropomorphism of AI impacts moral judgements such as blame \cite{malle_which_2016, waytz_mind_2014, young_autonomous_2019}, so it will in turn influence the desire for, and acceptance of punishment \cite{keijsers_teaching_2021, keijsers_whats_2021}. The reflexive relationship between anthropomorphism of AI and punishment judgements should be explored.

Punishment also communicates how much suffering we can justify. It has been noted that for punishment to truly act as a deterrent, we will need to program AI to be averse to punishment, which brings with it the problem of artificial suffering \cite{bryson__2017, harris_history_2022, metzinger_artificial_2021}.  Even where AI is not made to experience suffering, witnessing an apparently intelligent agent experience (perceived) harm may elicit negative emotions in humans. \citet{horstmann_robots_2018} found that participants only hesitated in turning off a social robot if the robot objected, stating it was afraid of the dark -- that it would suffer). We instinctively balk at a training paradigm that involved hitting a robot as punishment, even though it is virtual \cite{yu_behavior_2005}. AI need not truly experience suffering for humans to treat is as though it does, and this has implications for ensuring punishment is perceived as proportionate. \citet{harris_history_2022} quotes Christoph Bartneck as saying  that robot ``abuse'' is interesting because of how it reflects on us. Relatedly, \citet{darling_extending_2012} proposes extending legal protection to social robots to discourage similar abuse of human and non-human animals.  Punishment of AI may need to be seen as humane, regardless of what the technology itself experiences, because of the message that inhumane treatment spreads \cite{bryson_patiency_2018}. 

Relatedly, AI systems may be seen as proxies for different groups, and inhumane treatment of AI may encourage us to dehumanise actual humans \cite{bryson_patiency_2018}. Voice assistants, a common use example of AI, are typically gendered as female (despite limited attempts to counter this \cite{abercrombie_alexa_2021}) and their subservient role and casual responses to sexist abuse can reinforce gender stereotypes \cite{west_discriminating_2019}. Unjustified or overly cruel punishment of such systems could reinforce the patriarchal systems which enable violence towards women \cite{cercas_curry_metoo_2018}. 

\section{Open Questions}

We have drawn attention to key areas where future research is needed to understand how we might satisfy victims of AI harm through punishment of AI. This gives rise to many questions which we summarise under five themes. 
\begin{enumerate}
    \item \textbf{Public attitudes to legal punishment of AI. }
Intuitions amongst legal scholars differ as to whether direct punishment of AI will be satisfactory for the public. Our primary call to action is to investigate this. We acknowledge Lima and colleagues \cite{lima_explaining_2020,lima_conflict_2021} as having initiated such work, and call for expansion of their work to new scenarios and new testing paradigms. We ask ``How will use context and resulting harm impact expected forms of punishment?'' Further, ``Do the public equate deletion, re-assignment, loss of virtual assets etc with punishment?'' and ``Do these punishments achieve retributive, rehabilitative, deterrence and expressive goals, and how does this differ over proposed punishments?'' Finally, ``How might legal doctrine change to accommodate these expectations?'' 
\item \textbf{What people say, and what people do.}
Relatedly, there may be a conflict between the public’s reflective judgements on the punishment of AI \cite{lima_conflict_2021} and their impulsive behaviours \cite{ferdig_emotional_2004}. We call for work using different experimental paradigms to understand ``Will the public’s satisfaction with different forms of punishment reflect their ‘reflective’ or ‘impulsive’ judgements?''
\item \textbf{Learning from Robot(ic)s.}
HRI research can guide our initial exploration space, but crucially, evidence suggests we blame and punish disembodied AI differently \cite{bartneck_use_2006, malle_which_2016, malle_ai_2019}. We ask ``Do expectations for formal punishment of robots and AI differ?'' and ``Do social sanctions for robots and AI differ?''
\item \textbf{Blame: a Zero-Sum game?}
Several authors express the concern that AI will be blamed in place of developers, allowing their creators to evade punishment \cite{bigman_holding_2019, danaher_robots_2016, kneer_playing_2021} 
\cite[c.f.][]{shank_attributions_2018}, including in the so called ``court of public opinion''. We ask ``What conditions, if any, would lead to a shift in blame?'' 
\item \textbf{Anthropomorphism and Punishment. }
Building on topic (4), we ask ``Might anthropomorphic design be used to deliberately direct blame and change expectations for punishment?'', as anthropomorphism likely impacts how we punish AI \cite{keijsers_whats_2021, waytz_mind_2014}. Conversely, punishment may change how we anthropomorphise AI \cite{abbott_punishing_2024}. Thus we ask both ``How does perceived anthropomorphism of disembodied AI influence punishment judgements?'' and ``How does punishment of disembodied AI influence perceived anthropomorphism?''
\end{enumerate}

\section{Conclusion}

Given substantial evidence that the public are willing to ascribe blame to AI themselves when harm occurs, this gives rise to the question of how the public might like to see AI punished. The mechanics of punishing AI have been addressed in legal scholarship, but as yet been subject to limited empirical investigation with regards to public acceptance. We call for a comprehensive research programme to understand public expectations and answer the question of how victims of AI harm might be satisfied, lest we create a ``satisfaction gap'' whereby punishment fails to achieve one of its primary goals. Punishment of AI would change not only legal doctrine, but also perhaps AI design, governance, and our beliefs about AI and ourselves: cross-disciplinary scholarship is needed to explore these immense implications.

\section{Acknowledgements}
This work benefitted greatly from feedback by SJ Bennett, Nina Markl and Burkhard Schafer. Eddie L. Ungless is supported by the UKRI Centre for Doctoral Training in Natural Language Processing, funded by the UKRI (grant EP/S022481/1) and the University of Edinburgh, School of Informatics.

\bibliographystyle{ACM-Reference-Format}
\bibliography{references}


\end{document}